\documentclass[preprint,12pt]{aastex}
\usepackage{psfig}
\newcommand{\beq}{\begin{equation}}
\newcommand{\eeq}{\end{equation}}
\newcommand{\bea}{\begin{eqnarray}}
\newcommand{\eea}{\end{eqnarray}}
\newcommand{\bean}{\begin{eqnarray*}}
\newcommand{\eean}{\end{eqnarray*}}
\newcommand{\ba}{\begin{array}}
\newcommand{\ea}{\end{array}}
\newcommand{\bml}{\begin{mathletters}}
\newcommand{\eml}{\end{mathletters}}
\newcommand{\rem}[1]{{ }}
\bibpunct[,]{(}{)}{;}{a}{}{,}
\begin{document}
\title{Evidence for the Black Hole Event Horizon\footnote{George
Darwin Lecture presented at the Royal Astronomical Society, London,
December 13, 2002.  
}}

\author{Ramesh Narayan}

\affil{Department of Astronomy, Harvard University,
Harvard-Smithsonian Center for Astrophysics, 60 Garden
Street, Cambridge, MA 02138, U.S.A.}

\begin{abstract}

Astronomers have discovered many candidate black holes in X-ray
binaries and in the nuclei of galaxies.  The candidate objects are too
massive to be neutron stars, and for this reason they are considered
to be black holes.  While the evidence based on mass is certainly
strong, there is no proof yet that any of the objects possesses the
defining characteristic of a black hole, namely an event horizon.
Type I X-ray bursts, which are the result of thermonuclear explosions
when gas accretes onto the surface of a compact star, may provide
important evidence in this regard.  Type I bursts are commonly
observed in accreting neutron stars, which have surfaces, but have
never been seen in accreting black hole candidates.  It is argued that
the lack of bursts in black hole candidates is compelling evidence
that these objects do not have surfaces.  The objects must therefore
possess event horizons.

\end{abstract}

\section{Introduction}

The story of astrophysical black holes begins with a simple question:
How does a star, or any other gravitating object, hold itself up
against its own self-gravity?  In the case of the Sun, the answer is
easy.  The Sun is hot because of thermonuclear reactions in its
interior, and the resulting thermal pressure counteracts the
compressive action of gravity.  The same is true of all the stars we
see shining in the night sky.  When a star runs out of nuclear fuel
and dies, it must find other ways to fight gravity.  Dead stars with
masses up to the Chandrasekhar limit, $M_{\rm Ch}=1.4M_\odot$, become
white dwarfs, where electron degeneracy supplies the necessary
pressure.  Above the Chandrasekhar limit, and up to a second mass
limit, $M_{\rm NS,max}\sim2-3M_\odot$, dead stars become neutron
stars, where neutron degeneracy pressure holds them up.  But that,
according to conventional physics, is the end of the road.  If a dead
star has a mass $M>M_{\rm NS,max}$, there is no known force that can
hold the star up.  What we have then is a black hole, one of the most
extraordinary concepts in physics.

A black hole represents the ultimate victory of gravity, where all the
mass in the object collapses down to a ``singularity'', a true
geometrical point (at least within classical physics).  The object has
no material surface.  Instead, surrounding the singularity is an
``event horizon'', which plays the role of a virtual surface.  The
event horizon is a one-way membrane through which matter and energy
can fall in from the outside, but nothing, not even light, can escape
from within.  The region inside the horizon is thus causally cut off
from the outside world.  In a real sense, the horizon serves as an
effective surface, even though there is no actual material there.  For
a non-spinning black hole of mass $M$, the radius of the horizon is
given by Schwarzschild's result, $R_S=2.95(M/M_\odot)$ km.  Spinning
black holes have a somewhat smaller radius for the same mass.

Since Nature almost certainly makes dead stars with $M>M_{\rm
NS,max}$, black holes must exist in our Galaxy and in other galaxies.
Finding these black holes has been a central goal of high energy
astrophysicists for the last few decades, and indeed dozens of excellent
candidates have been discovered, many in the last ten years or so.
These discoveries are briefly summarized in \S2.  We then discuss in
\S\S3-5 the main theme of this Lecture: How can we confirm that the
black hole candidates discovered by astrophysics are truly black
holes?  Specifically, do the objects possess the defining
characteristic of a black hole, namely an event horizon?

\section{Astrophysical Black Holes}

Many excellent black hole candidates have been discovered in a class
of objects called X-ray binaries.  These are double stars in which a
compact primary star, either a neutron star or a black hole, accretes
mass from a normal secondary companion star (see Fig. 1), and radiates
the accretion luminosity in X-ray, ultraviolet and optical radiation.

A particular class of X-ray binaries, called soft X-ray transients or
SXTs for short, has turned out to be especially helpful in the hunt
for black holes.  In these binaries, the mass accretion rate $\dot M$,
and consequently the accretion luminosity $L_{\rm acc}$, vary with
time.  For most of the time, an SXT is in a very low luminosity state
with $L_{\rm acc}\sim10^{-6}-10^{-8}L_{\rm Edd}$, where $L_{\rm
Edd}=1.25\times10^{38}(M/M_\odot) ~{\rm erg\,s^{-1}}$ is the
Eddington luminosity and $M$ is the mass of the accreting star.  Every
once in a while, however, the system goes into an accretion outburst
and becomes very bright, with a luminosity almost equal to $L_{\rm
Edd}$.  It is during the accretion outburst that the binary is usually
discovered.  After the outburst, the luminosity slowly declines over a
period of several months, and the binary reverts to quiescence until
the next accretion outburst.

Whereas the optical light from an SXT during outburst comes mostly
from the accretion disk, the emission during quiescence is dominated
by the secondary star.  Therefore, in quiescence, it is possible to
observe the absorption lines in the stellar spectrum and use the
Doppler shifts on these to measure the orbital motion of the
secondary.  From the orbital period $P_{\rm orb}$ of the binary and
the semi-amplitude $K_s$ of the secondary's line-of-sight velocity,
one can then obtain the mass function
\begin{equation}
f(M) \equiv {P_{\rm orb}K_s^3 \over 2\pi G} = {\sin^3 i \over
(1+M_s/M_p)^2}M_p,
\end{equation}
where $M_s$ and $M_p$ are the masses of the secondary and the compact
primary, respectively, $i$ is the inclination angle of the binary
orbit, and the final relation on the right is derived by applying
Newton's laws of motion to a binary system bound by gravity in a
circular orbit.  Regardless of the values of $M_s$ and $i$, one sees
that $f(M)$ is a strict lower limit on $M_p$.

McClintock \& Remillard (1986) showed that the SXT A0620--00 has a
mass function close to $3M_\odot$, above the likely maximum mass
$M_{\rm NS,max}$ of a neutron star.  A0620--00 was the first truly
bona fide black hole candidate, and the sensational discovery of this
object opened up a new avenue in the hunt for black holes.  In the
years since, a dozen or more additional black hole candidates have
been discovered in SXTs, all identified as having mass functions $f(M)
\gtrsim M_{\rm NS,max}$.  Special mention should be made of the SXT
V404 Cyg (Casares, Charles \& Naylor 1992) which for many years was
the champion among black hole candidates, with the largest known mass
function of $6.1M_\odot$.  Clearly, the compact stars in A0620--00,
V404 Cyg, and the other SXTs with large mass functions, are all
excellent black hole candidates.  In a few SXTs and other kinds of
X-ray binaries, $f(M)$ itself is not very large, but there are
independent estimates of $i$ and $M_s$ which, combined with the
measured $f(M)$, indicate that $M_p > M_{\rm NS,max}$.  These objects
are also very good black hole candidates, though clearly not as good
as the systems with large mass functions.  Narayan, Garcia \&
McClintock (2002) list the best black hole SXTs as of 2001.

The typical black hole mass in SXTs is $\sim5-15M_\odot$.  In addition
to these stellar-mass black holes, supermassive black holes have been
identified in the nuclei of galaxies.  At the center of our own Milky
Way Galaxy, for instance, the object Sagittarius A* has been very
convincingly shown to be dark and compact, and to have a mass $\sim
4\times10^6M_\odot$ (see the remarkable observations described in
Sch\"odel et al. 2002 and Ghez et al. 2003).  Another excellent
candidate is in the nucleus of the galaxy NGC 4258 (Miyoshi et
al. 1995).  A couple of dozen additional candidates have been confimed
in the nuclei of other nearby galaxies (Gebhardt et al. 2003, and
references therein), and today it is generally agreed that virtually
every galaxy in the universe has a supermassive dark compact mass in
its nucleus.

\section{Is This Enough?}

As the above summary indicates, astrophysicists have certainly
discovered many compact stars that are too massive to be neutron
stars.  Can we therefore claim victory in the search for black holes?
In the opinion of the speaker, it would be premature to do so.

It is true that physics, specifically General Relativity combined with
our knowledge of the properties of matter up to nuclear density, tells
us that a compact star with $M>M_{\rm NS,max}$ cannot be a neutron
star and must therefore be a black hole.  However, what makes a black
hole unique --- indeed the reason why astronomers spend enormous
resources searching for it, and the public is so fascinated by it ---
is not that it is ``not a neutron star'' but that it possesses an
event horizon.  It is the event horizon that makes the black hole so
special.  Therefore, before claiming victory, it would be prudent to
look for independent evidence that the black hole candidates
identified so far on the basis of their masses actually do have event
horizons.  The speaker and his collaborators, Dr. Michael Garcia,
Dr. Jeremy Heyl and Dr. Jeffrey McClintock, have searched for such
evidence during the last several years.

What would constitute plausible evidence?  In brief, we need an
observed phenomenon --- or lack of it --- that is a unique signature
of an event horizon.  Ideally, we should compare black hole candidates
to a control sample of objects, say neutron stars, that are known to
have surfaces.  We should show that some observable characteristic is
distinctly different in the two classes of objects, and that the
difference is consistent with the notion that one class (black hole
candidates) has event horizons and the other class (neutron stars) has
surfaces.  Furthermore, the difference should not have any other
plausible explanation.  Of course, for this line of argument to work,
we need a working definition of what is plausible and what is not.
This tends to be somewhat subjective, though scientists by and large
agree on the matter.

X-ray binaries are particularly good for such investigations since
some X-ray binaries contain black hole candidates and some contain
neutron stars.  It is thus possible to find two well-matched sets of
binaries, one with black hole candidates that we think may have event
horizons, and the other with neutron stars that we know for sure have
surfaces.  The speaker and his collaborators pioneered such
comparisons of matched black hole and neutron star X-ray binareis, and
showed in an early study that quiescent black hole SXTs are very much
dimmer than quiescent neutron star SXTs (Narayan, Garcia \& McClintock
1997).  The latest observations are summarized in Fig. 2, taken from
McClintock et al. (2003).  As explained in Garcia et al. (2001) and
Narayan et al. (2002), the large luminosity difference --- a factor of
100 to 1000 in Eddington units --- is natural if black hole candidates
have event horizons.  Various alternative explanations have been
advanced in an effort to get around this argument, but many of these
explanations appear to be ruled out by subsequent observations
(Narayan et al. 2002).

Following our initial work, other signatures have been identified that
distinguish black hole and neutron star binaries.  Sunyaev \&
Revnivtsev (2000) showed that neutron star systems have strong
variability up to frequencies $\sim 1$ kHz, presumably from boundary
layer radiation close to the stellar surface, whereas black hole
systems have a significant decline in flux variations above $\sim
10-50$ Hz, presumably because they have no surface and hence do not
possess boundary layers.  In another study, Done \& Gierlinsky (2003)
considered the luminosities and spectra of black hole and neutron star
binaries and showed that there are large differences between the two
classes.  They suggested that the differences arise because black hole
candidates have event horizons while neutron stars do not.

We have recently begun a detailed study of Type I X-ray bursts in
X-ray binaries with a view to using the bursts to verify the
presence of event horizons in black hole candidates.  This is the
topic of the rest of the Lecture.

\section{Type I X-ray Bursts}

When gas accretes on the surface of a neutron star, it is compressed
by the strong surface gravity of the star.  As the gas sinks under the
weight of continued accretion, it becomes denser and hotter, until the
conditions are right for igniting thermonuclear reactions.  If the gas
consists largely of hydrogen and helium as it usually does, the
nuclear burning tends to be unstable (Hansen \& van Horn 1975, see
Lewin, Taam \& van Paradijs 1993, Bildsten 1998, for reviews).  The
instability causes the accreted layer of gas to burn its nuclear fuel
explosively within a very short time, leading to a burst of X-ray
emission.  After the fuel is consumed, the star reverts to its
accretion phase, in which fuel accumulates on the surface, until the
next thermonuclear instability is triggered.  The star thus undergoes
a semi-regular series of thermonuclear explosions.

Bursts of X-ray emission from X-ray binaries were first discovered by
Grindlay et al. (1976) and were immediately identified with
thermonuclear explosions on the surface of a neutron star.  These
explosions are known as Type I bursts (to distinguish them from a
different kind of burst, called Type II bursts, which is not
thermonuclear in origin).  In a typical Type I burst, the luminosity
of the neutron star increases to nearly the Eddington limit within
less than a second, and the flux then declines over a period of
seconds to tens of seconds.  The time interval between bursts is
usually several hours to perhaps a day or two.  Note that Type I bursts
are very distinct from the accretion outbursts described earlier.  The
latter are associated with changes in mass accretion rate and have
durations of months, not seconds, and recurrence times of many years,
not hours or days.  

Type I bursts are very common and have been seen in many neutron star
X-ray binaries.  However, it is a remarkable fact that no black hole
candidate in any X-ray binary has ever had a Type I burst.  In some
sense, it is obvious that these objects should not experience bursts.
A Type I burst requires a surface where matter is compressed and
heated until a thermonuclear instability is triggered.  A black hole
has no surface; matter simply falls in through the event horizon and
disappears.  Therefore, a black hole cannot have Type I bursts.  Since
black hole candidates are indeed observed not to have bursts, does it
then prove that they have event horizons?  The answer is,
unfortunately, ``No!''

An object that has a Type I burst must have a surface and therefore
cannot be a black hole.  This statement is uncontroversial.  However,
an object that does not have Type I bursts does not necessarily lack a
surface and therefore is not necessarily a black hole.  For instance,
there are bona fide neutron star systems, such as most X-ray pulsars,
that do not exhibit bursts even though they certainly have surfaces.
In order to use the lack of bursts as evidence for the presence of
event horizons, we need to carry out the following steps:

\begin{description}
\item{(i)} Develop a detailed theory of thermonuclear stability of
accreting gas on a compact star.

\item{(ii)} Use the theory to explain observations of Type I bursts in
neutron star systems and, in particular, explain why some neutron
stars have bursts and some do not.

\item{(iii)} Apply the theory to black hole candidates and demonstrate
that these systems would certainly exhibit bursts if they had
surfaces.

\item{(iv)} Show that the only plausible explanation for the lack of
bursts in black hole candidates is that the objects have event
horizons, and that all other explanations either are ruled out or are
very implausible.
\end{description}

Until we successfully accomplish these four stages, we cannot claim
that the lack of bursts in black hole binaries is strong evidence for
the event horizon.  Over the last two years, we have embarked on a
systematic attack on this problem.  As part of this work, we have
developed a theoretical framework for understanding the stability of
nuclear burning on compact stars (Narayan \& Heyl 2002, 2003).

We follow a star as it accretes gas of specified nuclear composition
(typically solar, corresponding to 70\% by mass of hydrogen, 28\% by
mass of helium, and 2\% by mass of heavier elements).  As the
accretion column grows, at each instant we solve for the
quasi-equilibrium configuration of the gas layer and obtain the
density, temperature and composition of the gas as a function of
depth.  We then carry out a linear stability analysis to determine if
the layer is stable to small perturbations.  If for some particular
column depth the layer is unstable, we say that the system undergoes a
Type I burst.  From the amount of unburnt fuel available in the layer
at the moment of instability, we estimate the net energy emitted in
the burst, and from the column density and the mass accretion rate, we
estimate the recurrence time between the bursts.  If the accreted
layer is stable for all column depths up to some very large value (say
$10^{13} ~{\rm g\,cm^{-2}}$), then we say that the system is able to
burn nuclear fuel stably and does not undergo bursts.  The main
innovation in our work is the application of a formal linear stability
analysis.  The approach is more rigorous than previous analyses, which
employed heuristic prescriptions to determine the stability of the
accreted gas.

\subsection{Application to Neutron Star X-ray Binaries}

Figure 3 summarizes the results obtained with this model for a neutron
star of mass $1.4M_\odot$ and various radii from $10^{0.2}R_S=6.5$ km
to $10^{0.6}R_S=16.4$ km.  The calculations correspond to a core
temperature of $T_{\rm core}=10^8$ K, which is approximately the
temperature expected for a core that cools via the modified URCA
process of neutrino emission.  This temperature is also close to the
upper limit to $T_{\rm core}$ in neutron star SXTs (see Narayan \&
Heyl 2002, 2003).  Figure 3 shows that, almost independent of radius,
accreting neutron stars are unstable to bursts at low accretion rates
$L_{\rm acc} \lesssim 0.25L_{\rm Edd}$ and stable at higher accretion
rates.  Observations indicate that bursts occur up to a luminosity of
about $0.25-0.3L_{\rm Edd}$ (van Paradijs et al. 1988; Cornelisse et
al. 2003; Tournear et al. 2003), and that systems brighter than this
limit have a much reduced bursting frequency.  The model is thus
consistent with the observations, with a tendency perhaps to
underestimate slightly the prevalence of burst behavior.

X-ray pulsars are generally found not to burst.  Although these
systems are usually not very luminous, nevertheless, because the
accretion is channeled onto the magnetic poles by a strong magnetic
field, the local mass accretion rate $\dot \Sigma$ (${\rm
g\,cm^{-2}s^{-1}}$) is generally above the Eddington rate.  Therefore,
the lack of bursts is consistent with the predictions of the model.
In the case of millisecond X-ray pulsars, however, the field is weak
and the accretion flow is only mildly channeled.  We would expect
these sources to burst, and indeed they do.

One could compare model predictions with more detailed observations of
burst recurrence times, burst durations, and other observables, as
discussed in Narayan \& Heyl (2003).  We do not describe the results
here, but merely note that the agreement overall is fairly good.  In
particular, some puzzling observations for bright systems with $L_{\rm
acc} > 0.1L_{\rm Edd}$, that previously had not been explained, appear
to be predicted satisfactorily by the new model.  For faint sources
with luminosities below about $0.03L_{\rm Edd}$ there appear to be
some discrepancies in the detailed model predictions, and for $L_{\rm
acc} < 0.01L_{\rm Edd}$ there are very few observations presently
available to check predictions.  This could be improved in the future.
In any case, we believe we have successfully accomplished stages (i)
and (ii) as listed in \S4, and we feel that the model is sufficiently
reliable to embark on stage (iii).

\subsection{Application to Black Hole X-ray Binaries}

Let us assume that black hole candidates in SXTs are not black holes,
but are some unusual kind of compact objects with surfaces.  Would
these objects have Type I X-ray bursts?  We can answer this question
using our burst model.  

First, we need to choose the core temperature and the stellar radius.
From the very low luminosities of black hole SXTs in quiescence, we
know that their core temperatures are no larger than $10^{7.5}$ K
(Narayan \& Heyl 2002).  We present results corresponding to this core
temperature, noting that the conclusions do not change very much for
lower temperatures.  We consider a range of likely radii for the
compact mass, from $(9/8)R_S$ (the smallest allowed radius, see
Shapiro \& Teukolsky 1983) up to about $2.8R_S$.

Figure 4 shows the results for a $10M_\odot$ object with a
hypothetical surface.  We see that such an object would have bursts
over a wide range of mass accretion rate.  That is, if SXT black hole
candidates have surfaces, they ought to produce bursts.  Consequently,
the absence of bursts is inconsistent with the presence of a surface.
This takes care of stage (iii) in \S4 and goes a long way towards
proving that black hole candidates must have event horizons.  What is
left to accomplish is the difficult stage (iv).

\section{Why Do Black Hole Candidates Not Have Type I Bursts?}

The lack of bursts in black hole candidates is certainly consistent
with the presence of event horizons in these objects.  This is not in
dispute.  But, is the presence of event horizons the only plausible
reason for the lack of bursts?  This is the critical question, and it
is where the argument becomes most difficult and subjective.  The
following discussion covers all the possible explanations for the lack
of bursts that the speaker considers plausible.

\subsection{Obvious Explanations}

Could black hole candidates be accreting the wrong kind of nuclear
fuel, e.g., something other than hydrogen and helium?  This is ruled
out because the companion stars that supply the gas are very normal,
with standard nuclear abundances.  Moreover, a strong hydrogen line is
seen in the optical spectra of all black hole SXT accretion disks,
confirming the presence of hydrogen in the accreting gas, and helium
lines are seen as well in many systems (Dr. J. E. McClintock, private
communication).

Could the core temperature be such that bursts are quenched?  Our
model calculations indicate that bursts are quenched only if the core
temperature is above several times $10^8$ K (Narayan \& Heyl 2002,
2003).  The very low X-ray luminosities of quiescent black hole SXTs
indicate that their cores are much cooler, $T_{\rm core} \lesssim
10^{7.5}$ K.  So this is not a promising explanation.

Could the burst recurrence times be extremely long, thereby making it
very difficult to detect bursts?  Figure 5 shows the dependence of the
recurrence time on the mass accretion rate for a typical black hole
candidate with a radius equal to $2R_S$.  The recurrence times are a
factor of a few longer than for neutron stars, as one might expect
since the accreting objects have larger masses, but the bursts are
correspondingly also more powerful by a factor of a few.  There is no
reason why these bursts should be missed by observations.  Figure 5
indicates that the bursts are most easily observed for $L_{\rm acc}
\gtrsim 0.1 L_{\rm Edd}$.  For lower accretion rates, the systems are
still burst-unstable (see Fig. 4), but the recurrence times are very
long and it is possible to miss these bursts.

\subsection{Wrong Accretion Rate?}

Figures 4 and 5 show that for certain accretion rates very close to
the Eddington rate bursts are either absent altogether or are very
rare.  Could it be that our black hole candidate systems spend all
their time in these particular regions of parameter space and that
this is why they do not burst?  In other words, is it just plain bad
luck that we have not seen bursts from these objects?  This
explanation can be eliminated easily.

Recall from \S2 that all our black hole candidates are found in
transient binary systems, so that each of them has experienced one or
more accretion outbursts during which its mass accretion rate varied
over a wide range.  At the start of an accretion outburst, a typical
system has a very low accretion rate, off beyond the left edge of
Fig. 4.  Then, as the outburst builds up, the accretion rate increases
and the system moves to the right almost to the right edge of the
figure, near the limit of Eddington accretion.  This journey typically
takes about a week.  Then, the accretion rate slowly declines over a
period of six months or so, during which the system moves slowly back
towards the left edge of the figure.  Finally, the system plunges back
into quiescence, beyond the left edge of the figure.

Figure 5 indicates that there is a favorable range of mass accretion
rate, between about 0.1 and 0.6 Eddington, for which bursts are
expected to occur quite frequently.  Although the particular
calculation shown corresponds to a specific choice of the radius, this
favorable region of bursting activity is present for all radii that we
have tried.  Furthermore, most black hole candidates have crossed this
region during their accretion outbursts, and each system has spent
about a month or so in the region.  During this time it should have
had several Type I bursts according to the calculations.  Therefore,
the fact that not a single source has experienced any Type I bursts is
very significant, and cannot be ``just plain bad luck.''

\subsection{Rotation and Magnetic Fields}

Could rapid rotation somehow eliminate bursts?  Rotation has the
effect of introducing a variation in the effective surface
gravitational acceleration $g$ as a function of latitude.  However,
even a maximally rotating compact star has only a factor of $\sim2$
variation in $g$ between the equator and the pole.  Such a modest
variation does not have a serious effect on the burst instability.

Rotation might influence the propagation of the burning front once the
instability has been triggered at a single point, and might thereby
have an effect on the development of the burst.  This is a complex
problem since the propagation of deflagration fronts is poorly
understood.  Although current calculations (e.g., Spitkowsky, Levin \&
Ushomirsky 2002), when applied to black hole candidates, seem to
indicate that bursts ought to develop readily in these systems, it
must still be considered an open question.

Neutron stars with strong magnetic fields do not burst.  The accretion
flows in these systems are channeled by the field, thereby increasing
the effective local mass accretion rate $\dot\Sigma$ to above the
Eddington rate.  Could black hole candidates have similarly
strong fields and thereby avoid bursts?  The possibility can be ruled
out immediately.  When accretion is channeled, the radiation comes out
in beams from the magnetic poles, causing the observed X-ray flux to
be modulated at the rotation period of the star.  This is, in
fact, the explanation for the modulations seen in X-ray pulsars.  If
black hole candidates have strong magnetic fields, they ought to
exhibit X-ray pulsations.  Since none of these objects has ever shown
coherent pulsations, we can rule out the magnetically channeled
accretion scenario.  In principle, if the magnetic axis is aligned
perfectly with the rotation axis there would be no observable
pulsation, but it requires such an unusual conspiracy in so many
objects that we consider it implausible.

\subsection{Exotic Stars}

Conventional physics tells us that black hole candidates cannot be
normal compact stars such as white dwarfs or neutron stars.  It is
therefore reasonable to suppose that, if they are not black holes,
then they must be exotic stars of some kind.  Could they be exotic
in such a manner as to prevent Type I bursts when they accrete gas?

The nature of matter deep in the core of a neutron star is uncertain
(Shapiro \& Teukolsky 1983; Glendenning 2000).  It might be normal
baryonic matter (mostly neutrons plus some protons and electrons), or
it might be a pion condensate, or it might be something more unusual
like Q-matter (Bahcall, Lynn \& Selipsky 1990), quark matter
(Glendenning 2000) or a color-locked superconducting state (Alford,
Rajagopal \& Wilczek 1998; Rapp et al. 1998).  Could black hole
candidates have cores made of some very exotic material, and could
this prevent bursts?  No, because bursts are very much a surface
phenomenon, restricted to low densities $\sim 10^6 ~{\rm g\,cm^{-3}}$.
The unusual kinds of matter that are invoked for neutron star
interiors typically occur at very high pressure, when the density is
above nuclear density ($\gtrsim10^{15} ~{\rm g\,cm^{-3}}$) and
certainly above the neutron drip point ($\sim 10^{11} ~{\rm
g\,cm^{-3}}$, despite the claim to the contrary by Abramowicz,
Kluzniak \& Lasota 2002 who seem to state that the color-locked
superconducting state can exist even at low densities).  Changes at
such high densities and such large depths have no effect on bursts at
the surface.

Could black hole candidates be strange stars or some other kind of
exotic stars where exotic matter extends all the way to zero pressure
at the surface?  Models of such stars have indeed been proposed
(Bodmer 1971; Witten 1984; see Glendenning 2000).  However, when the
stars accrete, it is believed that the newly added gas is supported
electrostatically as a separate surface layer of normal matter,
getting converted to exotic matter only at high density.  The
low-density layer would behave exactly as in our models and would
produce bursts just as in the calculations.

Is there any form of exotic matter that instantly converts infalling
baryonic matter into exotic matter even at densities below $10^6 ~{\rm
g\,cm^{-3}}$?  If a star were to be made of such material, there would
be no bursts since there would be no nuclei on the surface to undergo
thermonuclear burning.  To the knowledge of the speaker, no such
matter is presently known.

Finally, the black hole candidate may consist of some kind of dark
matter that does not interact with baryonic gas other than via gravity
(just like the dark matter in the universe).  We could then have a
$10M_\odot$ compact object that is completely ``porous'' to ordinary
matter.  Infalling gas would fall freely through the dark matter and
collect at the center.  (The speaker thanks Professor M. J. Rees for
suggesting this possibility.)  Would such an object have Type I
bursts?  To answer this question, we need to calculate the radius,
surface gravity and redshift of the baryonic gas ball that forms at
the center and then check whether or not the object would have Type I
bursts.  We have carried out some preliminary calculations which
suggest that the central gas sphere has characteristics not very
different from a neutron star and that the accreting gas does undergo
thermonuclear instability and Type I bursts.

\section{Conclusion}

The lack of Type I bursts in black hole SXTs is an important clue to
the nature of these compact stars.  Based on detailed models we
conclude that, if these objects possess surfaces, they should exhibit
widespread burst activity.  Why then do we not see Type I bursts in
black hole SXTs?

We have considered in \S5 a variety of explanations for the lack of
bursts and find that most can either be ruled out or can be ignored on
grounds of implausibility.  A couple of explanations are still viable.
One is that, once a thermonuclear instability is triggered, the
burning front, for some mysterious reason, is unable to propagate
quickly on the surface of a black hole candidate (\S5.3).  This
explanation, however, involves a considerable degree of special
pleading.  Burning fronts move quite rapidly on the surfaces of
neutron stars and white dwarfs, so why should they have trouble on
black hole candidates whose surface gravities lie in between?  The
other possibility is that black hole candidates are made of
non-interacting dark matter so that there is no hard surface where the
accreting gas can accumulate and undergo a thermonuclear instability
(\S5.4).  However, in this case, the gas would still accumulate at the
center and form a baryonic ball with a surface.  We have carried out
preliminary calculations of the bursting behavior of such objects and
our results suggest that bursts should be common.  Detailed results are
awaited.

Leaving aside the possibilities mentioned above, by far the most
plausible explanation for the lack of bursts in black hole candidates
is that the objects simply have no surfaces.  They must then have
event horizons.  Though not yet strong enough to qualify as proof,
this line of reasoning must surely be considered compelling evidence
for the reality of the event horizon.

\acknowledgements 
This work was supported in part by NASA grant NAG 5-10780 and NSF
grant AST 0307433.

\clearpage
\begin{figure}
\caption{Computer graphics image of an X-ray binary
(courtesy: Robert Hynes, 2002).  On the right is the donor secondary
star from which mass is transferred via a narrow stream to the compact
primary star on the left.  The gas goes into orbit around the primary,
and then spirals in via an accretion disk to accrete onto the compact
star at the center.  Some of the accreting gas is occasionally ejected
in twin jets, as shown.}
\end{figure}

\begin{figure}
\plotone{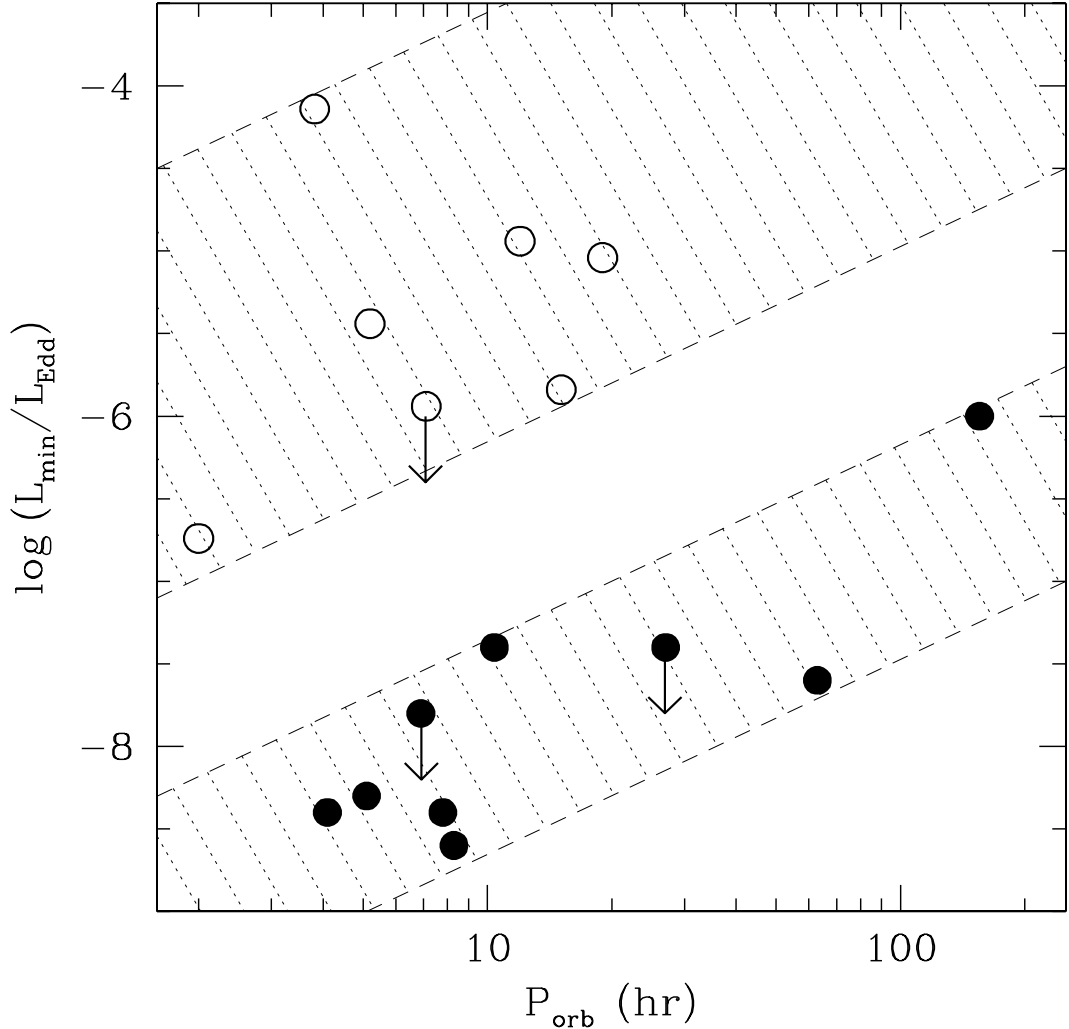}
\caption{Luminosities in quiescence, measured in
Eddington units, plotted against the orbital period, for 16 SXTs for
which data are available (taken from McClintock et al. 2003).  The
open circles correspond to SXTs with neutron star primaries and the
filled circles to systems with black hole candidates (identified by
mass).  The shaded bands have been drawn to emphasize that the two
classes of systems are well-separated on the plot.  Note that the
vertical axis is in logarithmic units, so that quiescent black hole
candidates are orders of magnitude dimmer than quiescent neutron
stars.  The large difference is arguably because black hole candidates
have event horizons whereas neutron stars have surfaces (Narayan,
Garcia \& McClintock 2002, and references therein).}
\end{figure}

\begin{figure}
\plotone{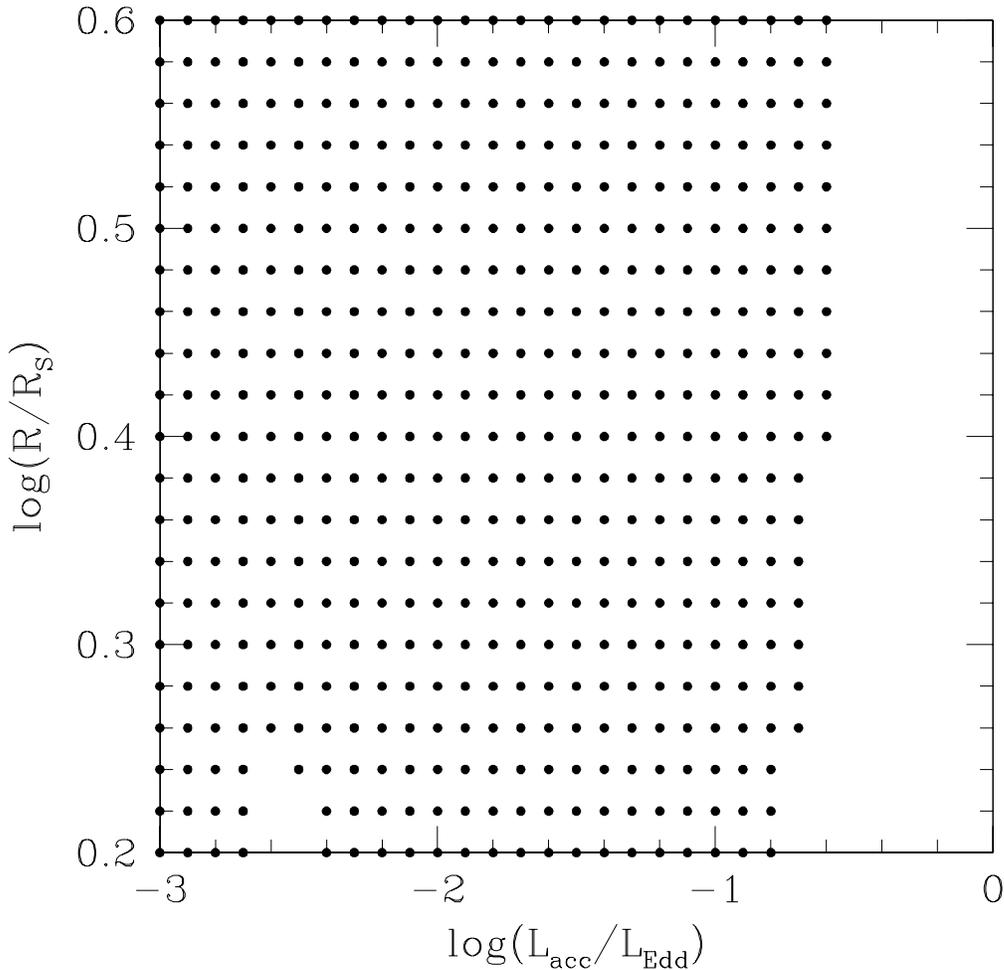}
\caption{Results on the thermonuclear stability of
accreting gas for a grid of neutron star models (Narayan \& Heyl
2003).  The neutron star is assumed to have a mass of $1.4M_\odot$ and
a core temperature of $10^8$ K.  The models correspond to a range of
stellar radii from $10^{0.2}R_S=6.5$ km to $10^{0.6}R_S=16.4$ km,
where $R_S$ is the Schwarzschild radius, and a range of mass accretion
rates from 0.001 Eddington to 1 Eddington.  The dots correspond to
models in which the accreting gas is thermonuclearly unstable and
produces Type I bursts.  The empty regions correspond to models in
which the gas accretes stably without bursts.}
\end{figure}

\begin{figure}
\plotone{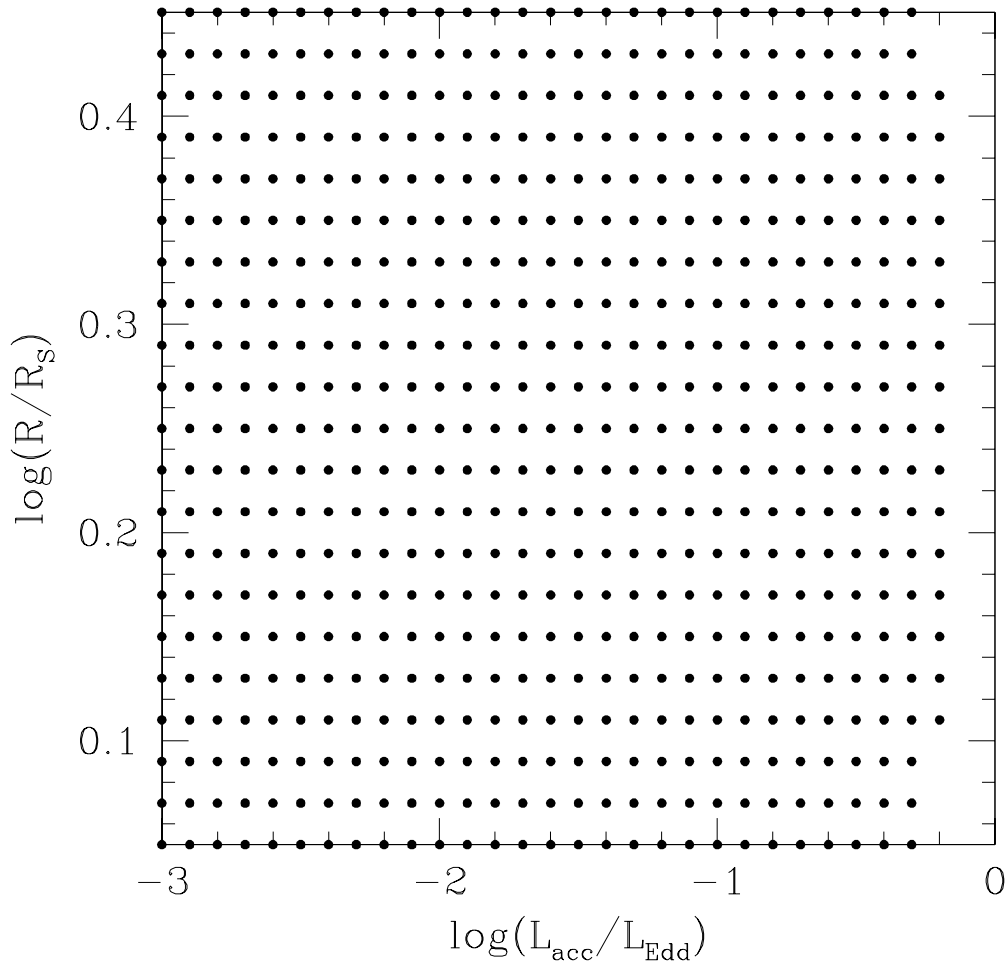}
\caption{Similar to Fig. 3, but for a hypothetical
black hole candidate with a surface.  The star is assumed to have a
mass of $10M_\odot$ and a core temperature of $10^{7.5}$ K.  The
models correspond to a range of radii from $(9/8)R_S=33$ km to about
$2.8R_S=83$ km, and a range of accretion rates from 0.001 Eddington to
1 Eddington.  The dots represent models in which this hypothetical
object would produce Type I bursts, and the empty regions to models
that are not expected to have bursts.}
\end{figure}

\begin{figure}
\plotone{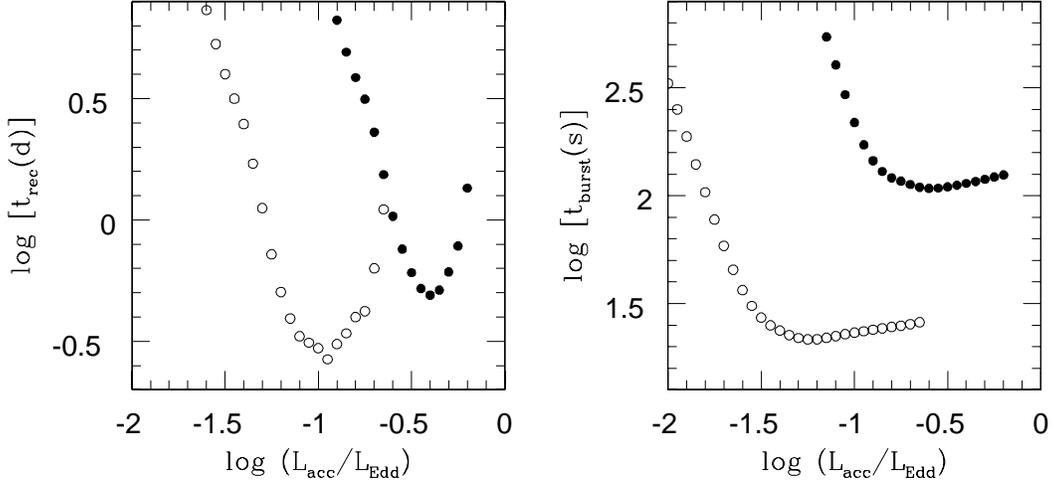}
\caption{Shows on the left the recurrence time in
days between bursts as a function of accretion rate for a $1.4M_\odot$
neutron star with a radius of $10^{0.4}R_S =10.4$ km (open circles)
and a hypothetical $10M_\odot$ black hole candidate with a surface at
a radius of $2R_S = 59$ km (filled circles).  Note that the black hole
candidate is expected to produce bursts with reasonably short
recurrence times for accretion rates $\sim0.1-0.6$ Eddington.  On the
right are the corresponding burst durations in seconds, assuming that
the burst energy comes out at the Eddington luminosity.}
\end{figure}

\end{document}